\documentclass[aps,groupedaddress,preprint,superscriptaddress,showpacs,showkeys]{revtex4-1}

\usepackage{amssymb}
\usepackage{amsmath}
\usepackage{graphicx}
\usepackage{epstopdf}
\usepackage{verbatim}

\begin{document}
\title{Low-noise kinetic inductance traveling-wave amplifier using three-wave mixing }

\author{M. R. Vissers}
\author{R. P. Erickson}
\author{H.-S. Ku}
\author{Leila Vale}
\author{Xian Wu}
\author{G. Hilton}
\author{D. P. Pappas}
\email[]{Electronic address: David.Pappas@NIST.gov}
\affiliation{National Institute of Standards and Technology, Boulder, CO 80305}

\date{\today}

\begin{abstract}
We have fabricated a wide-bandwidth, high dynamic range, low-noise cryogenic amplifier based on a superconducting kinetic inductance traveling-wave device. The device was made from NbTiN and consisted of a long, coplanar waveguide on a silicon chip.  By adding a DC current and an RF pump tone we are able to generate parametric amplification using three-wave mixing. The devices exhibit gain of more than 15 dB across an instantaneous bandwidth from 4 to 8 GHz. The total usable gain bandwidth, including both sides of the signal-idler gain region, is more than 6 GHz. The noise referred to the input of the  devices approaches the quantum limit, with less than 1 photon excess noise. Compared to similarly constructed four-wave mixing amplifiers, these devices operate with the RF pump at $\sim$20 dB lower power and at frequencies far from the signal. This will permit easier integration into large scale qubit and detector applications.
\end{abstract}


\maketitle

Recent advances in both quantum computing\cite{Wallraff-CQED} and detector \cite{Day-MKID} technologies have a established a need to measure large numbers of frequency-multiplexed microwave superconducting resonators at the quantum limit.  To address this need there has been a surge of research into superconducting amplifiers with wide frequency bandwidth, high dynamic range, and low noise. The prevalent current technologies for these amplifiers are based on highly nonlinear Josephson junctions (JJs) \cite{Bergreal,Beltran-JPA,Hatridge,Lafe-amp,  SLUG, Roch, Mutus-3WM}.  For currents well below the critical current, $I_C$, of the junctions, these devices typically have signal gain greater than 30 dB and added noise approaching the standard quantum limit, $SQL=\hbar\omega /2$, per unit bandwidth \cite{Caves}.  However, $I_C$ also sets the saturation power of the amplifier, limiting it to the order of the Josephson energy, $E_J=\hbar I_C / 2e$.  For typical parameters this is less than -110 dBm \cite{Mutus-3WM}. This sets the dynamic range of the amplifier, allowing it to measure only a few resonators simultaneously \cite{Mutus-IMJPA}. Future proposed applications of detectors \cite{Jonas-CCAT} and qubits \cite{Fowler-code} include orders of magnitude more devices that will need to be simultaneously read out. Furthermore, some JJ-based para-amps are embedded inside a resonant cavity. This design inherently limits the amplification bandwidth to the resonant cavity linewidth.  

New superconducting designs have been developed that aim to create broadband amplification using a four-wave mixing (4WM) traveling wave design\cite{Hansryd}. Devices based on either the nonlinear-kinetic inductance traveling waves (KITs)\cite{Eom,Clint-LTD} in NbTiN coplanar waveguides (CPWs) or Josephson junctions in a lumped LC transmission line \cite{JJ-TWPA,TWPA,White-JJ-TWPA} (typically referred to as TWPAs) have been demonstrated. Because they are not limited by the bandwidth of a resonant cavity, but rather the phase matching condition, these amplifiers can have bandwidths on the order of GHz. The primary difference between these technologies is that the KIT requires higher pump power than the TWPA in order to achieve the same nonlinearity. This gives the KIT intrinsically a higher dynamic range at the cost of requiring better filtering to separate the signal from the pump. 

In practice, traveling wave amplifiers require both momentum conservation, i.e., phase matching, and energy conservation for the pump, signal, and any idlers that are generated in the mixing process. For the KIT device momentum conservation can be attained by dispersion engineering of the coplanar waveguide (CPW) with periodic loadings that create a frequency gap. If the frequency gap is close to the pump, the dispersion-induced phase shift of the high-power pump can add to its self phase modulation, allowing it to match its cross phase modulation to other low power signals on the line. When this condition is met, parametric amplification can occur as the pump transfers power continuously to a signal and idler. Because the signal and idler frequencies lie on relatively linear portions of the band structure a wide bandwidth can be obtained. Additional band gaps can also be engineered to suppress higher harmonics of the pump \cite{Eom}. 

The KIT amplifier uses the unique properties of superconducting NbTiN, i.e., the low dissipation and high nonlinearity \cite{Eom,Clint-LTD,Erickson-comb}. In addition to the geometric inductance, $L_G$, superconductors exhibit a kinetic inductance, $L_K$, such that total inductance is $L_T=L_G+L_K$. Highly resistive superconductors such as NbTiN can have $L_K$ that exceeds 90$\%$ of the total inductance if they are made into structures with high critical current densities. Furthermore, $L_K$ is nonlinear in current, I, in the line, i.e., 

\begin{equation}	
\label{KI}
L_{K}(I)=L_0\left[1+{I^2\over{I_*^2}}\right]
\end{equation}

\noindent where the scaling parameter $I_*$ is of the order of the critical current and L$_0 \approx \hbar R_{s}/\pi\Delta$. The parameter $R_s$ is the normal sheet resistance and $\Delta$ is the BCS superconducting gap \cite{Mattis-Bardeen,Claasen,Gao-TiN,Annunziata}. For a device with an RF-only bias and signals, the current is $I=I_{RF}$, where $I_{RF}$ includes all possible RF components in the line. The low symmetry of the CPW geometry leads to a quadratic, Kerr-like dependence of $L_K$ as seen from energy conservation down the line, $E_{in}=E_{out}=L\times I^2 \propto I_{RF}^4$. This leads to the development of all possible RF 4WM products as the currents propagate down the line.

In this letter we report on the development of a three-wave mixing (3WM) \cite{Roch,Mutus-3WM} amplifier via the introduction of a DC bias, so $I\rightarrow I_{DC}+I_{RF}$. The nonlinear kinetic inductance from Eq.\ref{KI} therefore becomes 

\begin{equation}
L_K (I)=L_0 \left[1
+{I_{DC}^2\over I_*^2}
+2\left({I_{DC}I_{RF} \over I_*^2} \right)
+{I^2_{RF} \over I_*^2}
\right].
\end{equation} 

\noindent Again, from energy conservation, we now expect $E\propto I^3_{RF}+I^4_{RF}$, resulting in all possible 3WM as well as 4WM products at the output of the device. At the same time, we see that the kinetic inductance includes an extra $I^2_{DC}$ component, meaning that less RF input power is required to generate the same nonlinearity. Indeed, in our experiments with $I_{DC}>0$, we find the RF power needed to achieve a minimum 15 dB signal gain across the full bandwidth of 3WM is reduced by more than an order of magnitude when compared to 4WM. Furthermore, energy conservation requires that the pump, signal, and idler frequencies are related by $f_P = f_S + f_I$. Thus, when $f_S$ is varied the resulting signal gain will be symmetric about $f_P/2$ rather than $f_P$. This moves the pump frequency well away from the signal band, thereby allowing the pump to be easily filtered from the output. 

\begin{figure}
\includegraphics[width=400pt]{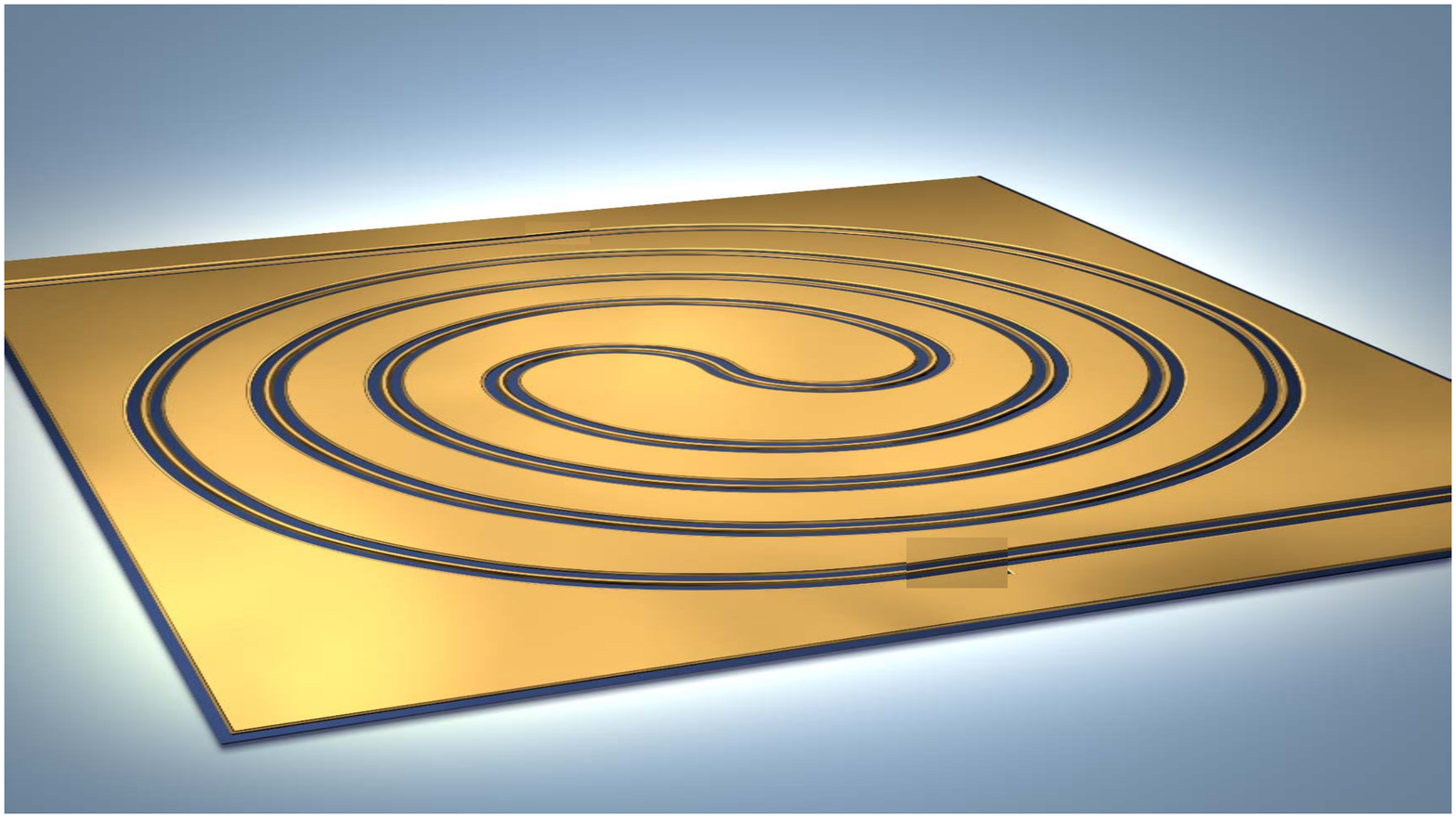}
\caption{\label{fig0} Schematic figure of the kinetic inductance traveling-wave (KIT) amplifier.}
\end{figure}

The KIT amplifiers were fabricated from a 15 nm film of NbTiN that was deposited by reactively co-sputtering from Nb and Ti targets in an Ar:N$_2$ atmosphere. Resonators fabricated out of this material showed very low loss, $\delta < 10^{-7}$, when measured as described in references \cite{JPL-TiN} and \cite{Vissers-TiN-APL}. The loss remained less than $10^{-5}$ even at $1\% \frac{\Delta L}{L}$ nonlinearity, and the KIT is typically operated well below these levels. To accommodate a relatively long line on a 2 cm $\times$ 2.2 cm Si chip the films were fabricated into spiral structures. A schematic view of the KIT chip is shown in Fig. [\ref{fig0}]. A CPW geometry was chosen to avoid dielectric loss and to simplify fabrication to a single step of optical lithography and a reactive ion etch. The CPW has a 3 $\mu$m center trace and a 3 $\mu$m gap.  The impedance of the CPW is $Z_{eff}\sim 200 \Omega$, as determined by simulation. The ground planes between the spiraled CPW center lines were about 100 $\mu$m wide and were coated with Au to suppress stripline modes. Tapered Klopfenstein impedance transformers on both ends were implemented to match to the $50 \Omega$ coaxial input/output cables. Similar to previous designs \cite{Eom,Clint-LTD}, the dispersion engineering was accomplished using periodic loadings created by widening the center strip of the CPW to 6$\mu$m at specific locations. 

Both low- and high-frequency devices were studied. The KITs reported here were 2 m long, corresponding to about 425 and 850 wavelengths for pump frequencies at about 8 and 16 GHz, respectively. A layer of Au was  deposited onto the back of the wafer to provide additional thermalization to the copper sample box \cite{Eom}. The box was then connectorized with SMA ports, transition boards were wire-bonded to the chip, and the assembly was cooled in a dilution refrigerator to 30 mK for measurements. Inside the refrigerator the RF tones, i.e. the pump and low power signal, were combined and then separated before and after the amplifier using diplexers at 30 mK. Similarly, $I_{DC}$ was applied using cryogenic-compatible bias T's. The output was then amplified with a high electron mobility transistor (HEMT) amplifier at the 4 K stage.

\begin{figure}
\includegraphics[width=400pt]{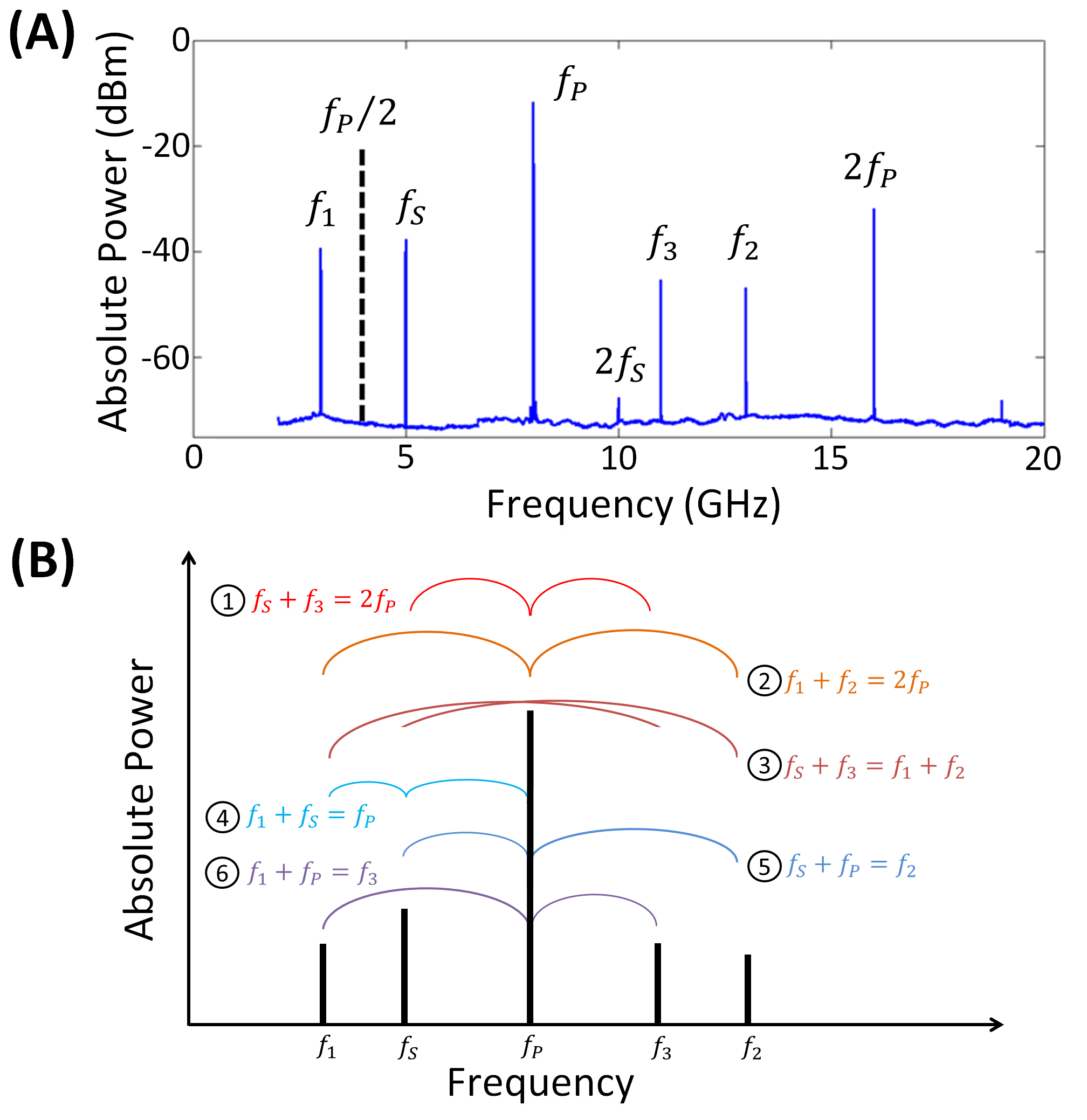}
\caption{\label{fig1} Measured and expected emission frequency spectrum of a KIT pumped with a 1 mA DC and -20 dBm RF bias. The input pump and signal frequencies are at $f_P=8$ and $f_S=5$ GHz, respectively. Panel (a)Experimental results - output signals are numbered, $f_1=3$ GHz, $f_2=13$ GHz, and $f_3=11$ GHz. These serve as idlers in first order and can be signals or idlers in second order, as shown in the lower panel. Panel (b)- Expected 3WM and 4WM mixing products to second order.}
\end{figure}

As a preliminary step in our investigations a spectrum analyzer was attached to the output of a NbTiN KIT. This first device was designed with a low frequency gap in order to observe the higher order mixing products. Figure \ref{fig1}(a) shows the measured tones generated when, for example, an 8 GHz pump with both DC and RF power is combined with a small, 5 GHz signal. On output, we observed three new tones, generated at $f_1=3$ GHz, $f_2=13$ GHz, and $f_3=11$ GHz, as labeled in the figure. These new tones are generated in the various possible 4WM and 3WM processes, as expected by energy conservation. Odd harmonics of $f_P$ and $f_S$ were blocked by the dispersion engineering but even harmonics are visible, resulting from this particular dispersion design. 

For comparison, Fig \ref{fig1}(b) shows the expected processes that result from energy conservation with the DC+RF pump combined with the signal. Processes are depicted out to second order where the generated idlers are allowed to act as new signals and mix with each other, the pump, and the signal. Specifically, there are three 4WM processes, (1) through (3), and three 3WM processes, (4) through (6), that can occur in this configuration. In addition, the processes numbered (3) and (6) are relatively weak in the presence of a strong, undepleted pump since they do not involve the pump directly. Similarly, while the 4WM process of (2) involves the pump, it also involves two idler products, and thus, as a secondary process, is relatively weak. This leaves as the most viable processes: (1) the degenerate 4WM process involving the idler of frequency $f_3$ with $f_S + f_3=2f_P$; (4) the 3WM processes, involving the idler of frequency $f_1$ with $f_1 + f_S=f_P$; and (5) the 3WM process involving the idler of frequency $f_2$ with $f_S + f_P=f_2$. All three of these remaining processes involve both the pump and signal, and hence, each can contribute to signal gain with a distinct and characteristic signature. In the cases of (1) and (4) the signal gain as a function of signal frequency $f_S$ is symmetric in character about $f_P$ and $f_P/2$, respectively, whereas in the case of (5) the signal gain is characteristically asymmetric.

\begin{figure}
\includegraphics[width=350pt, height=224pt]{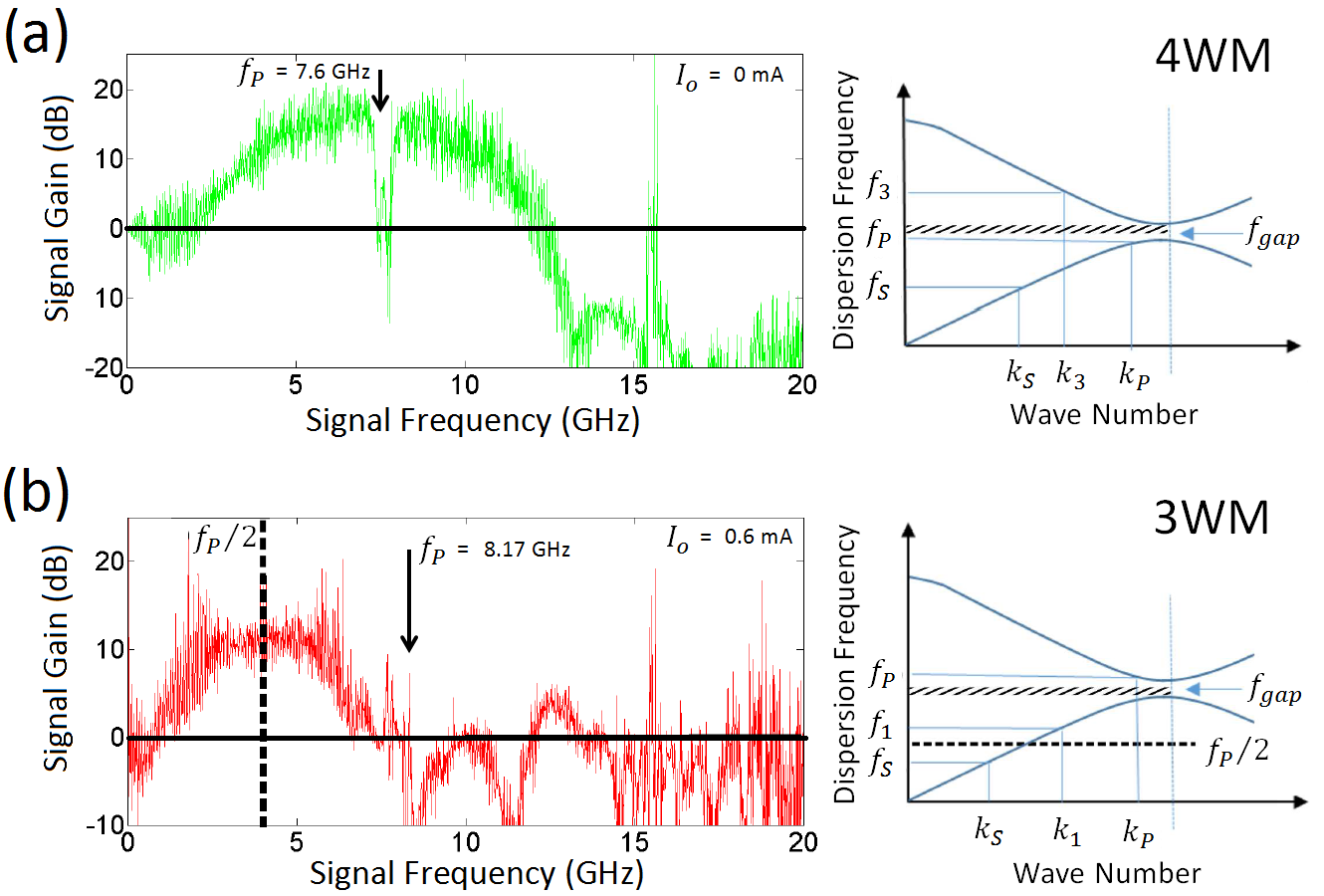}
\caption{\label{fig2} Measurements of signal gain as a function of signal frequency $f_S$ of the NbTiN KIT with first stop gap, $f_{gap}$, near 7.75 GHz. Panel (a) shows signal gain attributable to degenerate 4WM of type (1), as in Fig. \ref{fig1} (b), where no DC is applied. Here the device is pumped at $f_P=7.6$ GHz, which is below $f_{gap}$. At right of the measurements is a depiction of the matching of $f_P$, $f_S$, and $f_3$ to the lowest-lying dispersion curves of the device. Panel (b) shows signal gain attributable to 3WM of type (4), as in Fig. \ref{fig1} (b), with a DC of $I_DC=0.6$ mA applied. In this case the device is pumped at $f_P=8.17$ GHz, just above $f_{gap}$. At right of the measurements is an example depiction of the matching of $f_P$, $f_S$, and $f_1$ to the lowest-lying dispersion curves of the device.}
\end{figure}

We first tested the gain of the KIT used for Figure \ref{fig1}(a) with no DC bias, as shown in Figure \ref{fig2}(a). In this case, when the device was pumped just below $f_{gap}$, gain typical of 4WM is observed as described earlier by Eom, et al. \cite{Eom}. The signal gain is centered about $f_P$, characteristic of degenerate 4WM of type (1) of Fig \ref{fig1}(b). To the right of Figure \ref{fig2}(a) is a sketch of the dispersion-frequency bands showing the relative positions of $f_P$, $f_S$, and $f_3$. In this case the signal $f_S$ and 4WM idler $f_3$ straddle the gap--always one matched to the first band and the other to the second band, on either side of $f_P$. The notched region of frequency about $f_P$, corresponding to a loss of signal, occurs when either of $f_S$ or $f_3$ lies within the gap. Importantly, we find 4WM gain only for pump frequency just below the gap, and no signal gain is observed when the pump is just above the gap.

The device was then tested with a DC bias added, and the signal gain is shown in Figure \ref{fig2}(b). An important difference in this case is that continuous, broadband gain was only observed for pump frequencies just above the gap. In this case, the gain is observed symmetrically about the frequency $f_P/2$, and thus, is clearly attributable to the 3WM process of type (4) of Fig. \ref{fig1} (b). Some asymmetrical gain may be seen above $f_P$ that is relatively ineffectual and likely the result of the 3WM process of type (5). To the right of the plot the relative positions of $f_P$, $f_S$, and $f_1$ are shown on dispersion-frequency bands of the device, in accordance with the 3WM scenario of type (4). When the pump is moved to below the gap we find that signal gain is diminished with several frequency regions of narrow-band gain. This is likely the result of competition between the viable 3WM and 4WM processes. On the other hand, when the pump is above the gap the 4WM process may be inhibited by dynamics associated with the gap, thereby allowing the 3WM of the type (4) process to dominate over a wide bandwidth. A theory of parametric behavior accounting for the role of $f_{gap}$ is therefore warranted.

\begin{figure}
\includegraphics[width=350pt, height=278pt]{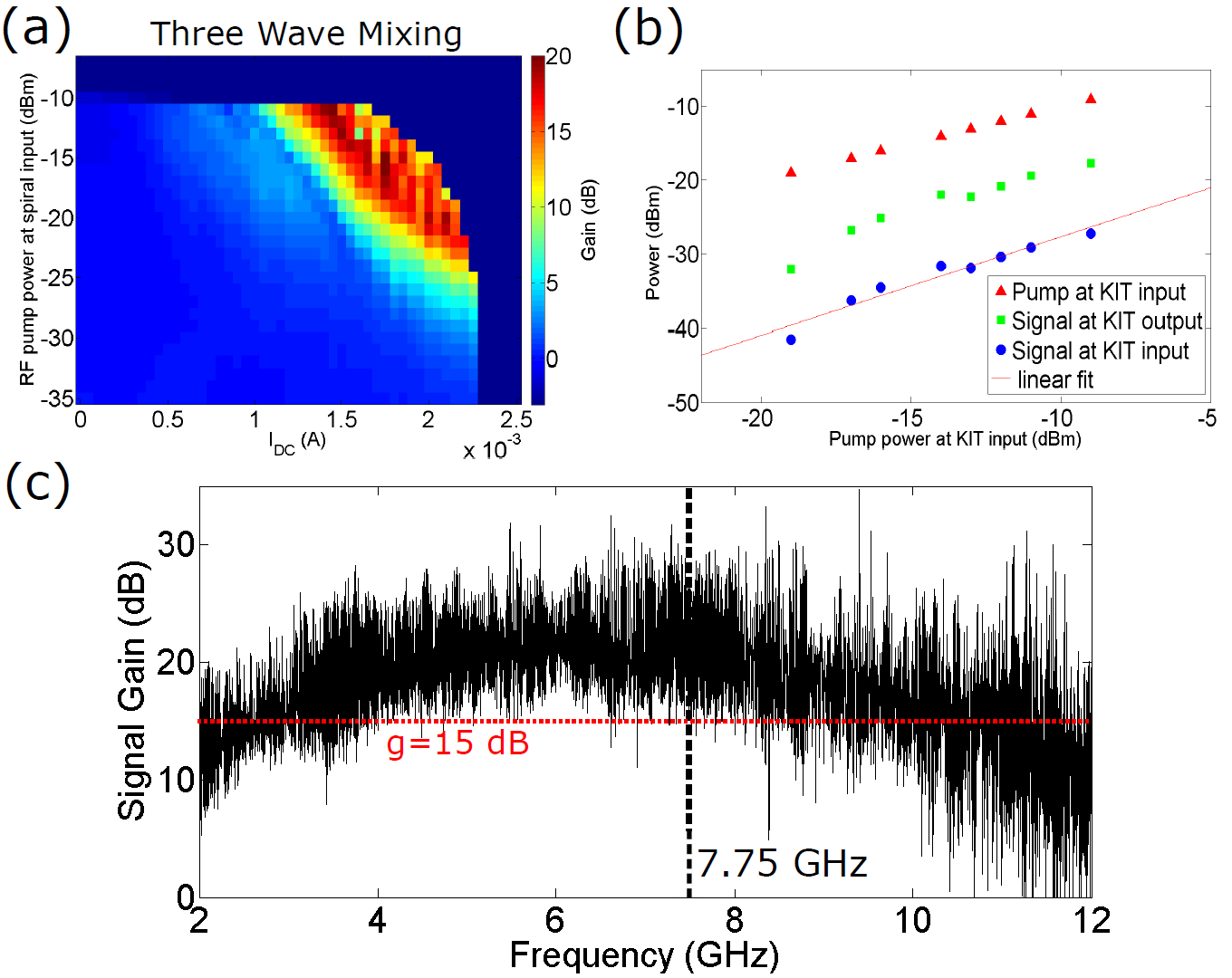}
\caption{\label{fig3} Sustainability and gain compression analysis of 3WM amplification for the KIT of first stop gap near 15.2 GHz, pumped at $f_P=15.5$ GHz. Panel (a) depicts in color scale the average signal gain of the KIT amplifier at 6 GHz $\pm$ 500 MHz for different RF pump powers and DC currents. Panel (b) shows the gain compression point of the amplifier at 10 dB gain for 8 different RF and DC bias points. Panel (c) illustrates the gain curve at a single bias point of panel (a) where the gain is above 15 dB (red dotted line) from 4 to 8 GHz, for all but a few data points.The black dashed line is at $f_P\over2$ }
\end{figure}

\begin{figure}
\includegraphics[width=350pt, height=350pt]{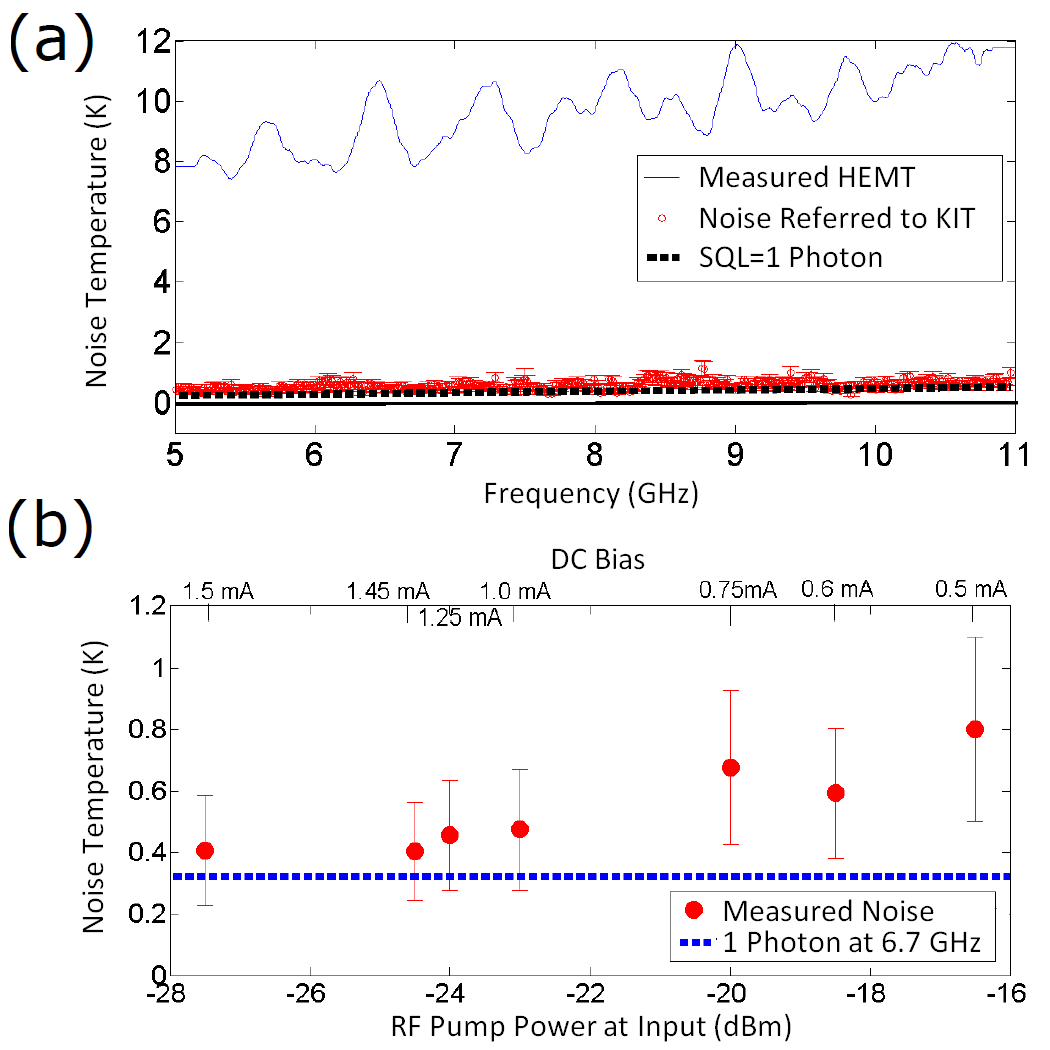}
\caption{\label{fig4} Noise analysis of a KIT pumped at 15.635 GHz. Panel (a) shows measured noise of the HEMT amplifier (blue), the noise referred to the KIT input (red), and the quantum limit (black), assumed to be $\hbar\omega$, as a function of signal frequency, with an RF input power of -23 dBm and an applied DC of 1.0 mA. The KIT noise is well below the HEMT and is on average 0.5 $\pm$0.3 photons above the quantum limit. Panel (b) depicts average noise referred to the input for 7 different RF-power/DC-bias points at the same pump frequency and gain. The DC biases are labeled on the top axis.}
\end{figure}

In order to measure the noise at typical frequencies used for qubit readouts, i.e. in the 4 to 8 GHz range, a 3WM chip was designed with the first dispersion gap at $\approx$15.2 GHz. Data from this device are shown in Figures \ref{fig3} and \ref{fig4}. This chip was pumped at with $f_P \approx 300$ MHz above the top edge of the gap and the 3WM gain was measured over a wide range of DC biases and RF powers. The resulting gain phase space is shown in Fig. \ref{fig3}(a), where value of the color scale is derived from the average gain over the frequency range from 6 to 7 GHz. With the DC bias current ranging from about 1 to 2.3 mA, more than 15 dB of gain was observed for RF pump powers from -10 to -30 dBm. These powers are 5 to 20 dB less than that required of 4WM devices. 

Another important parameter of these amplifiers is dynamic range. This will ultimately determine the total number of devices that can be read out by a single amplifier. The dynamic range is limited by the linearity of the device, and is typically defined by the 1 dB gain compression point. Figure \ref{fig3}(b) depicts the compression of the high frequency 3WM amplifier run at nominally 10 dB gain for several different RF and DC bias points at the same RF pump frequency. The 1 dB compression occurs above -45 dBm input signal power, about 5 dBm higher than that measured in the 4WM devices \cite{Eom}. Additionally, the 3WM amplifer permits the gain state to be acessible for a wide variety of DC and RF current combinations.  In general, bias points with greater RF power compress at higher signal powers. The data shows direct relationship between the RF power and gain compression, and we see that the gain saturates when the output signal:pump power$\approx$1:10.

To illustrate the broad-band nature of the amplifier gain, Figure \ref{fig3}(c) shows the signal gain curve at a single bias point from Fig.\ref{fig3}(a). Note that the gain is above 15 dB (red dotted line), from 4 to 8 GHz. However, one problem with the long spiral design is that it results in a split ground plane that can support parasitic resonances and slotline modes. The effect of this is that we observed significant ripple in both the $S_{21}$ and the signal gain, as can be seen in Fig. \ref{fig2} and Fig. \ref{fig3} (c). To mitigate this, additional Nb and Au layers were deposited onto the ground plane to reduce the sheet resistance and dampen resonances. Ultimately we plan to use either cross-over bridges or indium bump bonds with a flip-chip ground plane to eliminate these artifacts. We note that, in spite of this behavior, the current implementation still achieves gains of about 15 dB over a wide range. Moreover, we expect better phase matching when the ground plane is completed with either of the corrective methods mentioned.

Finally, the noise referred to the input of the high frequency 3WM KIT device and the HEMT was measured using a hot/cold load \cite{Hatridge, White-JJ-TWPA}. A switch just before the HEMT was used to toggle its input between a 50 $\Omega$ load at the 5.8 K 2nd stage and the 30 mK stage of the dilution refrigerator. Using this calibration for noise power, our HEMT noise temperature was measured to be between 8 to 10 K for our frequency region of interest, as shown in Fig. \ref{fig4}(a). To obtain the noise referred to the input of the KIT, four measurements of the noise power spectrum, $S_N$, were taken. These included measurements with the HEMT and KIT on and off as, 
\begin{equation}	
\label{noiseOnOff}
\begin{aligned}
S_N(\textrm{KIT off, Signal off}) &= G_{HEMT}A_{HEMT},\\
S_N(\textrm{KIT off, Signal on}) &= G_{HEMT}(S_{signal}+A_{HEMT}),\\
S_N(\textrm{KIT on, Signal off}) &= G_{HEMT}[G_{KIT}(A_{KIT} + S_{vac}) + A_{HEMT}],\\
S_N(\textrm{KIT on, Signal on}) &= G_{HEMT}[G_{KIT}(S_{signal}+A_{KIT} + S_{vac}) + A_{HEMT}],
\end{aligned}
\end{equation}

\noindent where $S_{signal}$ is the input signal, $G_{HEMT}$ and $G_{KIT}$ are the gain of the HEMT and KIT, respectively, and $A_{HEMT}$ and $A_{KIT}$ are their added noise. $S_{vac}$ is noise due to the vacuum, assumed to be $\hbar\omega / 2$. The measured noise gain, $G_N$, is found from dividing the two signal off traces and is $G_N=(G_{KIT}A_{KIT} + G_{KIT}S_{vac} + A_{HEMT})/A_{HEMT}$. Rearranging Eq. (\ref{noiseOnOff}), we have the added KIT noise 
\begin{equation}	\label{addedKitNoise}
A_{KIT} = (G_N - 1 ) \frac{A_{HEMT}}{G_{KIT}} -S_{vac} .
\end{equation}

From this, the noise referred to the KIT input, $A_{KIT} + S_{vac}$ was calculated over the full bandwidth, as shown in Fig. \ref{fig4} (a). The noise is close to the SQL ($\hbar\omega$), with only 0.5 $\pm$0.3 photons excess noise. Noise data taken at many bias points can be combined to show how the KIT noise changes with bias conditions. Figure \ref{fig4} (b) depicts average noise referred to the input for 7 different RF-power/DC-bias points at the same pump frequency and gain. The plot shows how the measured KIT noise increases at higher RF pump powers for the same gain. The DC biases are labeled on the top axis. Lower RF power is associated with the lower noise temperature. The higher RF power might induce greater chip temperatures or other forms of dissipation, i.e., chip heating, or incompletely thermalized electrons. Further optimization to run at lower RF powers or increased thermalization could result in lower measured noise. Another option is lower $T_C$ materials. Preliminary experiments with TiN have shown gain in the 3WM configuration at 10 dB lower RF pump power, consistent with the reduction in $T_C$ from 15 K to 5 K. 

We have shown that broadband, 15 dB gain with noise approaching the quantum limit can be achieved with a CPW NbTiN amplifier using 3WM. The 15 dB threshold is important because it allows the KIT to amplify signals that are close to the standard quantum limit at 30 mK, i.e. a typical temperature in a dilution refrigerated quantum computing circuit, to above the input noise of cryogenic HEMT amplifiers that are typically at 5-10 K.  Gain compression occurs at signal powers above -45 dBm, implying a very high dynamic range and the ability to frequency multiplex many devices. Compared to earlier 4WM configurations, the lower RF power requirement is expected to reduce thermal management issues and potential back action on to the measured devices\citep{Eom}. Also the large operational phase space permits optimization of the DC and RF bias points. Furthermore, the pump location at twice the frequency of the amplification region, instead of in the center as in the 4WM case, leads to not only a larger usable bandwidth but also reduces the requirements on the filters needed to separate the signal and pump paths, and eases integration with microwave devices. These improvements make the KIT amplifier with its high gain, high gain compression, large bandwidth, and nearly quantum-limited noise, an attractive candidate for reading out the large numbers of frequency multiplexed qubits or detectors that are required for the next generation application of superconducting device technologies.

\begin{acknowledgments}
This work was supported by the Army Research Office and the NIST Quantum Initiative.  RPE acknowledges grant 60NANB14D024 from the US Dept. of Commerce, NIST. We would also like to thank Lucas Heitzmann Gabrielli for his open-source library gdspy and Josh Mutus for useful discussions about three wave mixing in traveling wave devices as well as Jiansong Gao for additional support. This work is property of the US Government and not subject to copyright in the United States.
\end{acknowledgments}
\bibliography{3WM}

\pagebreak

\pagebreak

\pagebreak

\pagebreak

\pagebreak
\end{document}